\documentclass{article}
\usepackage{epsfig,amsmath,amssymb,natbib}

\begin{document}

\newcounter{saveeqn}
\newcommand{\alphaeqn}{\setcounter{saveeqn}{\value{equation}}%
\stepcounter{saveeqn}\setcounter{equation}{0}%
\renewcommand{\theequation}{\mbox{\arabic{section}.\arabic{saveeqn}\alph{equation}}}}
\newcommand{\reseteqn}{\setcounter{equation}{\value{saveeqn}}%
\renewcommand{\theequation}{\arabic{section}.\arabic{equation}}}

\begin{titlepage}

\title{The Role of Boundary Conditions in Solving Finite-Energy, Two-Body, Bound-State
Bethe-Salpeter Equations}
\author{G. B. Mainland \thanks{Supported by a grant from the Ohio Supercomputer Center}\\
Department of Physics\\
The Ohio State University at Newark\\
Newark, OH 43055, USA\\
\texttt{mainland@mps.ohio-state.edu}}
\maketitle

\begin{abstract}
The difficulties that typically prevent numerical solutions from being obtained to
finite-energy, two-body, bound-state Bethe-Salpeter equations can  often be overcome by
expanding solutions in terms of basis functions that obey the boundary conditions.  The
method discussed here for solving the Bethe-Salpeter equation requires only that the
equation can be Wick rotated and that the two angular variables associated with
rotations in three-dimensional space can be separated,  properties that are possessed
by many  Bethe-Salpeter equations including all two-body, bound-state Bethe-Salpeter
equations in the ladder approximation.  The efficacy of the method is demonstrated by
calculating finite-energy solutions to the partially-separated  Bethe-Salpeter equation
describing the Wick-Cutkosky model when the constituents do not have equal masses. 
\end{abstract}
\end{titlepage}

\setcounter{page}{2}

\section{Introduction}
\numberwithin{equation}{section}
\renewcommand{\theequation}{\arabic{section}.\arabic{equation}}

The Bethe-Salpeter equation \cite{Salpeter:51} is a covariant equation that, in some
sense, is a relativistic generalization of the Schr\"odinger equation  although it is
developed from relativistic quantum field theory rather than from relativistic quantum
mechanics.  One particularly noteworthy feature of the equation is that interactions
are retarded so that there is no action at a distance.  While the Bethe-Salpeter
equation is appropriate for studying properties of relativistic bound-state systems,
heretofore its use has been limited because, even numerically, the two-body,
bound-state equation has been exceedingly difficult to solve \cite{Mainland:99}.  For
this reason various approximations such as the Blankenbecler-Sugar approximation
\cite{Blankenbecler:66} or the instantaneous approximation
\cite{Salpeter:51, Salpeter:52} are often made that reduce the covariant equation in
four-dimensional space-time to a more tractable, approximately-covariant equation in
three dimensions.

If there are no external fields, the Bethe-Salpeter equation is rotationally invariant
in three-dimensional space so two angular variables can be separated.  Furthermore, at
least in the ladder approximation, the equation can be Wick rotated  (analytically
continued to Euclidean space) \cite{Wick:54}, which eliminates the singularity in the
kernel and makes the equation much easier to solve.  When a Bethe-Salpeter equation
has been partially separated and Wick-rotated, it is still an integral or differential
equation in two variables. The numerical method discussed here offers the possibility
of obtaining finite-energy solutions even when the equation is not completely
separable, which is usually the case, and and does not require that the masses of the
two bound quanta be equal.

With several exceptions, solutions to the two-body, bound-state Bethe-Salpeter
equation  have been obtained in the ladder approximation only when the equation is
completely separable or when the masses of the two bound quanta are equal. The 
Wick-Cutkosky model, which consists of two unequal-mass scalars interacting via a
massless scalar, is completely separable \cite{Wick:54, Cutkosky:54, Green:57, Seto:68,
Seto:68-erratum, Seto:69}, and the eigenvalue equation for the coupling constant can be
solved numerically \cite{Cutkosky:54, Linden:69a, Linden:69b}.  In the zero-energy
limit the Bethe-Salpeter equation is rotationally invariant in four-dimensional
space-time and is therefore separable.  Sometimes the completely separated equation has
been solved numerically.  For example, Brennan \cite{Brennan:75} obtained zero-energy,
bound-state solutions for two unequal-mass fermions interacting via a massive scalar. 
In the ladder approximation the author \cite{Mainland:03a, Mainland:03b} calculated
zero-energy solutions for a spin-0 and spin-1/2 constituent with masses that are not
equal and are bound by scalar electrodynamics.

Even if the equation is not completely separable, finite-energy, two-body,
bound-state solutions can occasionally be obtained if the masses of the two bound
quanta are equal. For example, Gammel and Menzel \cite{Gammel:73} determined the
bound-state solutions of two oppositely charged fermions that interact through
minimal electrodynamics.  Schwartz \cite{Schwartz:65} and  Nieuwenhuis and Tjon
\cite{Nieuwenhuis:96} determined bound-state solutions for two scalars that interact
via a third, massive scalar. When the two bound scalars have unequal masses,
finite-energy solutions were first obtained   by Kaufmann \cite{Kaufmann:69} and later
by Set\^{o} and Fukui \cite{Seto:93}, who reduce the Bethe-Salpeter equation to an
infinite system of integral equations in one variable that are solved numerically.  In
all cases, before the equations are solved,  they are Wick-rotated \cite{Wick:54} to
eliminate the singularity in the kernel.  

Because the energy appears in more than one place in the Bethe-Salpeter equation, a
Hamiltonian does not exist, and the equation is an eigenvalue equation for the
coupling constant instead of for the energy.  The equation is solved by specifying a
value for the energy and then, for the chosen value of energy, calculating values of
coupling constant that satisfy the equation. Although the coupling constants are real
in the Lagrangian,  there are apparently solutions to the Bethe-Salpeter equation with
complex values of the coupling constant.  While the Wick-Cutkosky model
\cite{Wick:54, Cutkosky:54}  has only real values for the coupling constant,
Kaufmann \cite{Kaufmann:69}  considered two scalars interacting with a third, massive
scalar and found complex values.  Also, for the same equation Set\^o and Fukui
\cite{Seto:93} found  that ``there exists a strong indication that complex eigenvalues 
appear$\dots$ .''  Here attention is restricted to solutions of the Bethe-Salpeter
equation  with real values of the coupling constants, which are the more interesting
physically.

Numerical solutions to the bound-state, Bethe-Salpeter equation are obtained in
five steps:  

1)  The singularity in the kernel is removed by a Wick rotation \cite{Wick:54}, which is
always possible in the ladder approximation,  and  is accomplished by making
the substitution $p_{0} \rightarrow ip_{0}$ while rotating the path of integration
$90^\circ$ counterclockwise in the complex  $p_0$-plane.  

2)  Two angular variables are separated, which is possible
because the Bethe-Salpeter equation is rotationally invariant in three-dimensional
space provided there are no external fields. The resulting  equation for the
Bethe-Salpeter ``wave function" $\Psi(ip_{0},p_{s})$ is an equation in the two
variables $p_0$ and $p_s \equiv |{\bf p}|$. In the ladder approximation a Wick-rotated,
partially-separated Bethe-Salpeter equation is of the form 
\begin{equation}
K(ip_0,p_s)\Psi (ip_0, p_s) = {g_1g_2 \over 4\pi} \int_{- \infty}^\infty \
\mathrm{d}q_0
\int_0^\infty
 \mathrm{d}q_s V(i p_0, p_s,i q_0, q_s)\Psi (iq_0, q_s). 
\end{equation}
The above equation actually represents $N_{EQ}$ equations, where $N_{EQ}=1$ if both
constituent quanta have spin zero and $N_{EQ} >1$ otherwise.  Thus, $K(ip_0,p_s)$ and
the kernel $V(i p_0, p_s,i q_0, q_s)$ are both $N_{EQ} \times N_{EQ}$ matrix
functions.  

3)  Zero-energy solutions are  calculated. In the zero-energy limit the
Bethe-Salpeter equation is invariant under rotations in four-dimensional
space-time, and is, therefore,  completely separable. Zero-energy solutions are
expanded in terms of basis functions that consist of the product of a set of basis
functions $\{g_i(|p|)\}$ that depend on the magnitude of the Euclidean four-momentum
$|p|=(p_0^2+p_s^2)^{1/2}$ and  hyperspherical harmonics in four-dimensional, Euclidean
space-time. To obtain solutions, each of the basis functions $g_i(|p|)$ must (very
nearly) obey the boundary conditions, which are readily calculated \cite{Mainland:03b}. 
Each basis function need not obey the boundary conditions exactly provided that a linear
combination of the basis functions yields a solution that does. 

4) Finite-energy solutions $\Psi(ip_{0},p_{s})$ are expanded in terms of a set of
basis functions \{$g_j(p_0, p_s)$\},  
\begin{equation}
\Psi (i p_0, p_s) = \sum_{j=1}^{N_B} c_j g_j (p_0, p_s).
\end{equation}
Two conditions are imposed on the basis functions: (a) The basis functions must (very
nearly) obey the boundary conditions. (b) A basis system must be chosen  that,
in the zero-energy limit, devolves to the basis system that yields zero-energy
solutions.  Knowledge of a basis system that yields zero-energy solutions provides
guidance in constructing a  more general basis system required to represent
finite-energy solutions.  

5) Finally, the partially separated Bethe-Salpeter equation
(1.1) is discretized by converting it into a generalized matrix eigenvalue equation for
the coupling constant. One additional condition is imposed on the
generalized matrix eigenvalue equation: In the zero-energy limit the 
generalized matrix eigenvalue equation that yields finite-energy solutions must devolve
to the   generalized matrix eigenvalue equation that yields zero-energy solutions. 
Discretization can be accomplished, for example, using the Rayleigh-Ritz-Galerkin method
\cite{Delves:74, Atkinson:76} or the method of orthogonal polynomials
\cite{Mainland:03a}. After expressing the solution $\Psi(ip_{0},p_{s})$ in terms of
basis functions,  both sides of (1.1) are multiplied  by
$f(p_0,p_s)\,g_i(p_0,p_s)^\dagger$ and then integrated  over the variables $p_0$ and
$p_s$.  The  function $f(p_0,p_s)$ may be omitted or may be chosen so that that the
matrices are symmetric or have some other desirable property.  The integral equation (1.1) has then  been converted into a generalized matrix
eigenvalue equation
\begin{equation} 
Kc = {g_1g_2 \over 4\pi} (V_H + V_{AH})c.
\end{equation}
In the above equation, $c$ is a column vector with the elements $c_j$ that are the
expansion coefficients for the wave function $\Psi (i p_0, p_s)$ in (1.2), and the
matrices $V_H$ and  $V_{AH}$ are Hermitian and anti-Hermitian, respectively.   Since
the Bethe-Salpeter wave function is expressed in terms of $N_B$ basis functions as
indicated in (1.2),  (1.3) is an $(N_{EQ} \times N_B) \times (N_{EQ} \times N_B)$
matrix equation. 

Because there is no obvious way to force the eigenvalues of (1.3) to be real, in
general it has been extremely difficult to construct a generalized matrix eigenvalue
equation that yields real values for $g_1g_2/4\pi$ that are solutions to (1.1)  A
sufficient condition for obtaining real eigenvalues of a generalized matrix eigenvalue
equation (1.3) is that $V_{AH}=0$, $K$ be Hermitian and either $K$ or $V_H$ be positive
definite. (See, for example, \cite{Hildebrand:65}.)  In (1.3),  $K$ is often 
Hermitian.  But if $K$ is also  positive definite, then $V_{AH}$ is usually non-zero,
and if  $V_{AH}$ is  zero, then neither $K$ nor $V_H$ is usually  positive definite.
And even if an eigenvalue of (1.3) happens to be real, especially when the basis
functions do not obey the boundary conditions, the eigenvalue typically is not an
eigenvalue of the Bethe-Salpeter equation (1.1). 

Solutions to some partially separated Bethe-Salpeter equations have been obtained when
the masses of the two constituents are equal because, in this case, the matrix  $K$ in
(1.3) is both Hermitian and positive definite, and the matrix $V_{AH}$ vanishes
because it is proportional to the difference of the masses of the two constituents. 
For example, for the equal-mass and zero-energy cases, the Bethe-Salpeter equation
describing the  Wick-Cutkosky model \cite{Wick:54, Cutkosky:54} can be
converted into a matrix equation of the form (1.3) where both $K$ and
$V_{H}$ are real, symmetric and positive-definite and $V_{AH}=0$ because it is
proportional to the mass difference of the two bound quanta as well as the energy
of the bound state \cite{Mainland:95}.

When a matrix eigenvalue equation is constructed such that the conditions discussed in
steps 1) - 5) are satisfied, all eigenvalues usually are not  real.  But real
eigenvalues are obtained, and almost all real eigenvalues are solutions of the original
Bethe-Salpeter equation.

To demonstrate the  techniques for solving a finite-energy, two-body, bound-state
Bethe-Salpeter equation  as well as  the effectiveness of the method, finite-energy
solutions are calculated for the partially separated Wick-Cutkosky model when the
constituents masses are unequal.  Although the equation is separable and the solutions
were originally calculated from a completely separated equation, the method used here
 only requires that the two angular variables associated with rotations in
three-dimensional space be separated. The advantage of demonstrating the technique with
the Wick-Cutkosky model is that the complications associated with higher spin are
avoided. Earlier the author suggested an alternative method \cite{Mainland:99} for
solving such equations and then used the method to obtain solutions to the the partially
separated Wick-Cutkosky model.  However, the solutions obtained here are more accurate,
the numerical method is less difficult to implement, is more widely applicable and is
more efficient.
 
\section{The Bethe-Salpeter Equation for the  Wick-Cutkosky Model}
\numberwithin{equation}{section}
\renewcommand{\theequation}{\arabic{section}.\arabic{equation}}

The Wick-Cutkosky model \cite{Wick:54, Cutkosky:54} consists of two scalars with
respective masses $m_1$ and $m_2$ that interact with a third massless scalar.  In the
ladder approximation, the Bethe-Salpeter equation that describes a bound state of the
two massive scalars is
\begin{align}  
\{(p^\mu + \xi K^\mu)(p_\mu + \xi K_\mu) -& m_1^2\}\{[p^\nu + (\xi-1)K^\nu][p_\nu +
(\xi-1)K_\nu]-m_2^2\}\chi_K(p)\nonumber\\
&={i\lambda \over \pi^2}  \int_{- \infty}^{\infty} {\mathrm{d}^4q \over
(p-q)^2+i\epsilon}\chi_K(q),
\end{align}  
where the notation is that of \cite{Bjorken:65}. The parameter $0 <\xi < 1$ in
the above equation is  associated with the definition of the center-of-mass variables,
and
$K^\mu$ is the four momentum of the bound state. After a Wick rotation \cite{Wick:54},
in the rest frame of the center of mass where
$K^\mu=(E,0,0,0)$, the Bethe-Salpeter equation takes the form, 
\begin{align}  
\{(ip^0 + \xi E)^2-{\bf p}^2 &- m_1^2\}\{[ip_0 + (\xi-1)E]^2-{\bf p}^2
-m_2^2\}\chi_E(ip_0,\bf p)\nonumber\\
&={\lambda \over \pi^2}  \int_{- \infty}^{\infty} {d^4q \over
(p-q)\cdot (p-q)}\chi_E(iq_0,\bf q),
\end{align} 
\noindent where $(p-q)\cdot (p-q) \equiv (p_0-q_0)^2+(p^i-q^i)(p^i-q^i)$ is the
Euclidean scalar product. 

Dimensionless variables are introduced by defining $m_1\equiv m(1+\Delta)$, $m_2\equiv
m(1-\Delta)$, dimensionless momentum $p'\equiv p/m$ and dimensionless energy $\epsilon
\equiv E/2m$.  When written in terms of dimensionless parameters, the above equation
becomes
\begin{align} 
\{(ip^0 + 2\xi \epsilon)^2-{\bf p}^2 &- (1+\Delta)^2\}
\{[ip_0 + 2(\xi-1)\epsilon]^2-{\bf p}^2
-(1-\Delta)^2\}\chi_E(ip_0,{\bf p})\nonumber\\
&={\lambda \over \pi^2 m^2}  \int_{- \infty}^{\infty} {d^4q \over
(p-q)\cdot (p-q)}\chi_E(iq_0,{\bf q}).
\end{align}
where primes have been omitted  since all momenta are now dimensionless.

For compactness of notation, it is convenient to write the coefficient of $\chi
{_E}(ip_0,{\bf p})$ on the left-hand side of (2.3) explicitly in terms of its real and
imaginary parts,
\begin{align}
 &\{(ip^0 + 2\xi \epsilon)^2 -{\bf p}^2 - (1+\Delta)^2\} \{[ip_0 + 2(\xi-1)
\epsilon]^2-{\bf p}^2-(1-\Delta)^2\}\nonumber\\ &\equiv D{_R} +iD{_I}.
\end{align}
From (2.4) it immediately follows that
\alphaeqn
\begin{align}
D{_R}&=[p_0^2 + {\bf p}^2 -4\xi^2\epsilon^2 + (1+\Delta)^2] 
[p_0^2 + {\bf p}^2
-4(1-\xi)^2\epsilon^2 + (1-\Delta)^2]\nonumber\\
&+16 \xi(1-\xi)\epsilon^2p_0^2,\\
D{_I}&=4 \epsilon p_0 \{-\xi [p_0^2 + {\bf p}^2 -4(1-\xi)^2\epsilon^2 +
(1-\Delta)^2]\nonumber\\ 
&+(1-\xi)[p_0^2 + {\bf p}^2 -4\xi^2\epsilon^2 +
(1+\Delta)^2]\}.
\end{align}
\reseteqn
Because $D_I$ vanishes both in the zero-energy limit, $\epsilon=0$,  and, if
$\xi=1/2$, when the two constituents have equal masses, $\Delta =0$, it is relatively
easy to obtain solutions in these two limits.

Since the coupling constant $\lambda$ is real in the Lagrangian, the physically
interesting values of $\lambda$ are real. Actually, for the Wick-Cutkosky model all
eigenvalues are real \cite{Cutkosky:54, Linden:69a, Linden:69b}  although for other
Bethe-Salpeter equations,  solutions may exist for complex values of the
coupling constant as discussed previously \cite{Kaufmann:69, Seto:93}. 

Writing $\chi_E (ip_0,\bf p)$ in terms of real and imaginary parts,
\begin{equation}
\chi_E(ip_0,{\bf p}) \equiv \chi_R(p_0,{\bf p})+i\chi_I(p_0,{\bf p}),
\end{equation}
and noting that the real and imaginary parts of (2.3) must vanish independently, yields
the following two equations:
\alphaeqn 
\begin{align}
&D{_R}\chi{_R}(p_0,{\bf p})-D_I\chi_I(p_0,{\bf p})={\lambda \over
\pi^2 m^2}  
\int_{- \infty}^{\infty} {d^4q \over(p-q)\cdot (p-q)}\chi_R(q_0,{\bf q}),\\
&D_I\chi_R(p_0,{\bf p})+D_R\chi_I(p_0,{\bf p})={\lambda \over \pi^2 m^2}  
\int_{- \infty}^{\infty} {d^4q \over(p-q)\cdot (p-q)}\chi_I(q_0,{\bf q}).
\end{align}
\reseteqn 
Adding (2.7a) and (2.7b),
\begin{eqnarray}
D{_R}[\chi{_R}(p_0,{\bf p})+\chi_I(p_0,{\bf p})]+D_I[\chi_R(p_0,{\bf p})
-\chi_I(p_0,{\bf
p})]\nonumber\\
={\lambda \over \pi^2 m^2}  
\int_{- \infty}^{\infty} {d^4q \over(p-q)\cdot (p-q)}
[\chi_R(q_0,{\bf q})+\chi_I(q_0,{\bf q})],
\end{eqnarray}

From (2.3) it  immediately follows that if
$\chi_E(ip_0,{\bf p})$ is a solution, then $\chi_E ^*(-ip_0,{\bf p})$ is a solution
with the same eigenvalue. Thus, without loss of generality it is possible to choose
\begin{equation}
\chi_E(ip_0,{\bf p})=\chi_E ^*(-ip_0,{\bf p}).
\end{equation}
Taking the complex conjugate of (2.9),
\begin{equation}
\chi_E^*(ip_0,{\bf p})=\chi_E(-ip_0,{\bf p}).
\end{equation}
Therefore, the real and imaginary parts of the solution can be chosen, respectively,
to be even and odd functions of $p_0$.

Defining
\begin{equation}
\psi(p_0,{\bf p}) \equiv \chi{_R}(p_0,{\bf p})+\chi_I(p_0,{\bf p}),
\end{equation}
it immediately follows that
\begin{equation}
\psi(-p_0,{\bf p}) \equiv \chi{_R}(p_0,{\bf p})-\chi_I(p_0,{\bf p}).
\end{equation}
Consequently (2.8) can be rewritten as
\begin{equation}
D{_R}\psi(p_0,{\bf p})+D_I\psi(-p_0,{\bf p})
={\lambda \over \pi^2 m^2}  
\int_{- \infty}^{\infty} {d^4q \over(p-q)\cdot (p-q)}\psi(q_0,{\bf q}),
\end{equation}
which is in a form that is convenient to solve numerically.

\section{ Numerical Solutions to the Partially Separated, Wick-Cutkosky Model}
\numberwithin{equation}{section}
\renewcommand{\theequation}{\arabic{section}.\arabic{equation}}

Since (2.13) is manifestly invariant under rotations in three-dimensional space, the
angular dependence associated with such rotations separates. The wave function
$\psi(p_0,{\bf p})$ can be written in the form
\begin{equation}
\psi(p_0,{\bf p})=F(p_0,|{\bf p}|) Y^{\ell}_m(\theta,\phi),
\end{equation}
where $Y^{\ell}_m(\theta,\phi)$ is a spherical harmonic.    The integration over the
two angular variables on the right-hand  side of (2.13) can be performed analytically
using Hecke's theorem \cite{Hecke:18}, and the angular dependence of the
solution  then separates.  Unfortunately, with this approach the remaining kernel is
an associated Legendre function containing  a logarithmic singularity that is
difficult to integrate over numerically \cite{Mainland:01}.  Furthermore, the two
remaining  integrations on the right-hand side of (2.13) must be performed numerically.

An easier method for solving (2.13) is achieved by first rewriting the equation in
terms of spherical coordinates in four-dimensional, Euclidean space-time
\cite{Levy:50, Nieuwenhuis:96}:
\begin{align}  
	p^0		& = 	|p| \cos \theta_1 \nonumber &
	p_z		& = 	|p| \sin \theta_1 \cos \theta_2 \nonumber \\
	p_x		& = 	|p| \sin \theta_1 \sin \theta_2 \sin \phi  &
	p_y		& = 	|p| \sin \theta_1 \sin \theta_2 \cos \phi
\end{align} 
The four-momentum $q^\mu$ is written similarly in terms of primed angles.

The solution $\psi(p_0,{\bf p})$ is then expressed as a series expansion in terms of
hyperspherical harmonics $P^{(2)}_{k,\ell}(cos\hspace{.1
cm}\theta_1)Y^{\ell}_m(\theta_2,\phi)$ in four-dimensional, Euclidean space-time.  
Defining
$z
\equiv cos\hspace{.1 cm}\theta_1$,  the spherical function $P^{(2)}_{k,\ell}(z)$ is
given by \cite{Mainland:86}  
\begin{equation}
P^{(2)}_{k,\ell}(z)=(1-z^2)^{\ell/2} {d^{\ell}  \over \mathrm{d}z^{\ell}} C^1_k(z),
\end{equation}
where $C^1_k(z)$ is a Gegenbauer polynomial.  Now $C^1_k(z)$ is an even or odd
function of $z$ if the integer $k$ is even or odd, respectively.  From (3.3) it then
immediately follows that
$P^{(2)}_{k,\ell}$ is an even or odd function of $cos\hspace{.1 cm}\theta_1$ if
$k-\ell$ is respectively, an even or odd integer. Recalling that $\chi{_R}(p_0,{\bf
p})$ and $\chi_I(p_0,{\bf p})$ are, respectively, even and odd functions of $p_0$,
implying that they are also, respectively, even and odd functions of $cos\,\theta_1$,
zero-energy solutions can be obtained from expansions of the form
\alphaeqn 
\begin{align}
&\chi{_R}(p_0,{\bf p})_{\mathrm {zero-energy}}=\sum_{n=1}^{N_p}
g_n\,G_n(|p|)P^{(2)}_{\ell+i,\ell}(cos\,
\theta_1)Y^{\ell}_m(\theta_2,\phi),\\
&\chi{_I}(p_0,{\bf p})_{\mathrm {zero-energy}}=\sum_{n=1}^{N_p}
g_n\,G_n(|p|)P^{(2)}_{\ell+1+i,\ell}(cos\,
\theta_1)Y^{\ell}_m(\theta_2,\phi).
\end{align}
\reseteqn 
In the above expansions, $g_n$ is an expansion coefficient, the index $i=0,2,\dots$ is
an even integer and \{$G_n(|p|)$\} is a set of basis functions, each of which (very
nearly) obeys the boundary conditions. 

A generalization of the zero-energy basis system (3.4) that is suitable for calculating
finite-energy solutions is 
\alphaeqn 
\begin{align}
&\chi{_R}(p_0,{\bf p})=\sum_{n=1}^{N_p}
\hspace{.3cm}\sum_{k=\ell,\ell
+2,...}^{K_{max}}g_{n,k}G_n(|p|)P^{(2)}_{k,\ell}(cos\,
\theta_1)Y^{\ell}_m(\theta_2,\phi),\\
&\chi{_I}(p_0,{\bf p})=\sum_{n=1}^{N_p}
\hspace{.3cm}\sum_{k=\ell +1,\ell
+3,...}^{K_{max}}g_{n,k}G_n(|p|)P^{(2)}_{k,\ell}(cos\,
\theta_1)Y^{\ell}_m(\theta_2,\phi).
\end{align}
\reseteqn 
The values of the index $k$ in (3.5) are chosen so that $\chi{_R}(p_0,{\bf
p})$ and $\chi_I(p_0,{\bf p})$ are, respectively, even and odd functions  of
$cos\,\theta_1$. In the above expansions, $g_{n,k}$ is an expansion coefficient and
\{$G_n(|p|)$\} is a set of basis functions, each of which (very nearly) obeys the
boundary conditions and will  be specified later. Recalling that
$\psi(p_0,{\bf p}) = \chi{_R}(p_0,{\bf p})+\chi_I(p_0,{\bf p})$,  the expansion for
$\psi(p_0,{\bf p})$ immediately follows from (3.5),
\begin{eqnarray}
\psi(p_0,{\bf p})=\sum_{n=1}^{N_p}
\hspace{.3cm}\sum_{k=\ell,\ell
+1,...}^{K_{max}}g_{n,k}G_n(|p|)P^{(2)}_{k,\ell}(cos\,
\theta_1)Y^{\ell}_m(\theta_2,\phi).
\end{eqnarray}

In the zero-energy limit, the angular dependence of the solution separates and only $\chi{_R}(p_0,{\bf
p})$ or $\chi_I(p_0,{\bf p})$ is nonzero.
Thus, zero-energy solutions can be obtained from (3.6) by choosing one value of the
parameter $k=\ell, \ell+1,\dots $, with each different value of $k$ yielding
different solutions.  As a consequence, in the zero-energy limit the basis system
(3.6) devolves to a suitable  basis system for obtaining zero-energy solutions. 

There are three advantages to seeking solutions of the form (3.6) instead of (3.1): 1) 
After using Hecke's theorem \cite{Hecke:18} to perform the three angular
integrations analytically, the remaining kernel does not contain a logarithmic
singularity. 2) In (2.13) only one integration must be performed numerically instead of
two.    3)  The basis functions have the correct angular dependence for zero-energy
solutions so that fewer angular terms are required  to obtain accurate, finite-energy
solutions when the states are tightly bound.

Substituting (3.6) into (2.13), 
\begin{align}
\sum_{n=1}^{N_p}
\hspace{.3cm}&\sum_{k=\ell,\ell
+1,...}^{K_{max}}g_{n,k}G_n(|p|)[D{_R}P^{(2)}_{k,\ell}(cos\hspace{.1 cm}
\theta_1)+D_IP^{(2)}_{k,\ell}(-cos\hspace{.1 cm}
\theta_1)]Y^{\ell}_m(\theta_2,\phi)\nonumber\\
&={\lambda \over \pi^2 m^2}  
\sum_{n=1}^{N_p}
\hspace{.3cm}\sum_{k=\ell,\ell
+1,...}^{K_{max}}g_{n,k}\int_{0}^{\infty}\mathrm{d}|q|\hspace{.1 cm}|q|^3\,G_n(|q|)
\nonumber\\ 
&\times\int {\mathrm{d}\Omega'_{(3)}
\over(p^2+q^2-2pq\,cos\,\Theta)}P^{(2)}_{k,\ell}(cos'\hspace{.1 cm}
\theta_1)Y^{\ell}_m(\theta'_2,\phi '),
\end{align}
where $\Theta$ is the angle between the four-vectors $p$ and $q$. Using Hecke's
theorem \cite{Hecke:18} to perform the angular integration (All necessary formulas are
in the appendix of Ref. \cite{Mainland:86}.),
\begin{align}
&\sum_{n=1}^{N_p}
\hspace{.3cm}\sum_{k=\ell,\ell
+1,\dots}^{K_{max}}g_{n,k}G_n(|p|)[D{_R}P^{(2)}_{k,\ell}(cos\hspace{.1 cm}
\theta_1)+D_IP^{(2)}_{k,\ell}(-cos\hspace{.1 cm}
\theta_1)]Y^{\ell}_m(\theta_2,\phi)\nonumber\\
&={\lambda \over \pi^2 m^2} 
\sum_{n=1}^{N_p}
\hspace{.3cm}\sum_{k=\ell,\ell
+1,...}^{K_{max}}g_{n,k}\int_{0}^{\infty}\mathrm{d}|q|\,|q|^3\,G_n(|q|)
\Lambda^{(2)}_k(|p|,|q|) P^{(2)}_{k,\ell}(cos\hspace{.1 cm}
\theta_1)Y^{\ell}_m(\theta_2,\phi).
\end{align}
The function $\Lambda^{(2)}_k(|p|,|q|)$ is \cite{Hecke:18, Mainland:86},
\begin{equation}
\Lambda^{(2)}_k(|p|,|q|)=\left \{
\begin{array}{c}
\frac{2\,\pi^2}{|p||q|(k+1)}\left(\frac{|q|}{|p|}\right)^{k+1}\hspace{.3 cm}
\mbox{if}\hspace{.5 cm} |q|\leq |p|\\
\\
\frac{2\,\pi^2}{|p||q|(k+1)}\left(\frac{|p|}{|q|}\right)^{k+1}\hspace{.3
cm}\mbox{if}\hspace{.5 cm} |p|\leq |q|.\\
\end{array}
\right. 
\end{equation} 
The dependence on the angular variables $\theta_2$ and $\phi$ separates as it must.

To determine the boundary conditions, the parameters $g_0$ and $g_\infty$ must be
calculated that satisfy
\alphaeqn
\begin{align}  
& G_n(|q|) _{\stackrel{\displaystyle\longrightarrow}  {|q| \rightarrow 0}}
|q|^{g_0},\\ 
&G_n(|q|) 
_{\stackrel{\displaystyle\longrightarrow}  {|q| \rightarrow \infty}}
 |q|^{-g_\infty}.
\end{align}
\reseteqn
Once the asymptotic behavior of  integrals of the form
\begin{equation}
I(p) =\int_0^\infty \mathrm{d}|q|\,
|q|^n\, G(|q|)\, \Lambda^{(2)}_k(|p|,|q|),	
\end{equation}
which appears in (3.8), are determined, the boundary conditions are readily calculated.
Specifically, the parameters
$i_0$ and $i_\infty$ must first be calculated that, respectively, satisfy 
\alphaeqn
\begin{align}  
 &I(|p|) _{\stackrel{\displaystyle\longrightarrow}  {|p| \rightarrow 0}}
|p|^{i_0},\\
&I(|p|) _{\stackrel{\displaystyle\longrightarrow}  {|p| \rightarrow \infty}}
   |p|^{-i_\infty}.
\end{align}
\reseteqn
There are two possible values for the parameter $i_0$ in (3.12a)\cite{Mainland:03b}:
\alphaeqn
\begin{align}  
&\mathrm{Solution\; IA}: & &i_0=k, & &-n+k+1\leq g_0\\
&&&&&n-k-1<g_\infty\nonumber\\ 
&\mathrm {Solution\; IB}: & &i_0=g_0+n-1,&&-n-k-1<g_0<-n+k+1 \\
&&&&&n-k-1<g_\infty\nonumber 
\end{align}
\reseteqn
Similarly, there are two possible values for the parameter $i_\infty$ in (3.12b)
\cite{Mainland:03b}:
\alphaeqn
\begin{align} 
&\mathrm{Solution\; IIA}: & &i_\infty=k+2, & & -n-k-1<g_0\\
&&&&&n+k+1\leq g_\infty\nonumber\\ 
&\mathrm {Solution\; IIB}: & &i_\infty=g_\infty-n+1, & &-n-k-1<g_0\\
&&&&&n-k-1<g_\infty \leq n+k+1\nonumber  
\end{align}
\reseteqn

Using the fact that as $|p|\rightarrow0$, $D_R \rightarrow$ constant, $D_I \rightarrow
|p|$ and substituting  (3.10a) into (3.8), for Solution IA  it follows that
\begin{equation} 
|p|^{g_0}+|p|\,|p|^{g_0}\sim|p|^k.
\end{equation}
Thus, $g_0=k$. Because the smallest value of k=$\ell$,
\begin{equation} 
G_n(|p|) _{\stackrel{\displaystyle\longrightarrow}  {|p| \rightarrow 0}}
|p|^{\ell}.  
\end{equation}
As can be readily checked, there are no other solutions for $g_0$.  Similarly, at
large $|p|$, the only solution is $g_\infty=k+6$ so
\begin{equation} 
G_n(|p|) _{\stackrel{\displaystyle\longrightarrow}  {p \rightarrow \infty}}
{1 \over |p|^{\ell +6}}.
\end{equation}

The knot structure is as follows: There are $N_p$ cubic splines in the expansion (3.6)
and $N_p+4$ momentum knots $T_p(i)$. To determine the momentum knots,
$N_p$ Chebyshev points $x_p(i)$ are calculated on the interval
$-1<x_p(i)<1$,
\begin{equation} 
x_p(i)=-\mathrm{cos}\frac{(2i-1)\,\pi}{2\,N_p},\hspace{1.0
cm}i=1,2,\dots,N_p.
\end{equation}
The momentum knots $T_p(i+4)$ are then given by
\begin{equation}
T_p(i + 4) = C^\prime \sqrt {{1 + x_p(i) \over 1 - x_p(i)}} +
C^{\prime \prime}, \hspace{1.0 cm} i = 1, 2, \dots, N_p.
\end{equation}
The constant $C^\prime$ is chosen by trial and error to approximately minimize the
lowest zero-energy eigenvalue, and the constant $C^{\prime \prime}$ is chosen so that
the first knot on the positive $|p|$-axis is not too close to $|p| = 0$.  The values
$C^\prime = 1.0$ and $C^{\prime \prime} = 0.01$ were satisfactory. A knot is placed at
the origin, $T_p(4) = 0$, and the three knots $T_p(1), T_p(2)$ and $
T_p(3)$ are placed on the ``negative'' $|p|$ axis to allow maximum freedom in
constructing the solution from splines near $|p| = 0$.  The three knots on the
``negative'' $|p|$ axis are mirror images (about the origin) of the first three knots
in (3.19). With this choice of knots, the first three splines are finite at the
origin,  creating sufficient freedom to construct solutions from splines near
$|p|=0$

Angular knots are  chosen on the $z$ axis, where $z=\mathrm{cos}\,\theta_1$, so
that numerical integrations can be carried out over cos$\,\theta_1$. Defining
$N_\theta \equiv K_{max}-\ell+1$, which is the number of hyperspherical harmonics in
the expansion (3.6) of the solution, arbitrarily, but in analogy with splines, the
number of angular knots $T_z$ is chosen to be $N_\theta +4$. The angular knots
$T_z(1)=-1$,
$T_z(N_\theta +4)=1$ and the remaining knots are the Chebyshev points
\begin{equation} 
T_\theta(i+1)=-\mathrm{cos}\frac{(2i-1)\,\pi}{2\,(N_\theta+2)},\hspace{1.0
cm}i=1,2,\dots,N_\theta+2.
\end{equation}
So that  the basis functions $G_n(|p|)$ asymptotically vanish as indicated in (3.16)
and (3.17), $G_n(|p|)$ is chosen as follows:
\begin{equation} 
G_n(|p|) =  \frac {|p|^{\ell}} { a+|p|^{2\ell +5}} B_n(|p|) \equiv
{\mathcal G}_{\ell}(|p|)B_n(|p|),
\end{equation}
where ${\mathcal G}_{\ell}(|p|)$ is a convergence function, ``$a$'' is a constant and
$B_n(|p|)$ is a cubic spline \cite{Boor:78}. At small $|p|$, $ G_n(|p|) \sim 
|p|^{\ell}\,B_n(|p|)$.  Since the splines are also functions of $|p|$, at small $|p|$
the individual basis functions $G_n(|p|)$ do not exactly obey the boundary condition
(3.16). However,  as $|p| \rightarrow 0$, for each solution   the sum
of the first three splines in the expansion (3.6)  approaches a
constant so that each solution satisfies the boundary condition  exactly.
At large $|p|$, since all splines vanish,  the convergence function is chosen to
vanish  as $1/|p|^{\ell +5}$, which is one power of $|p|$ slower than the rate in
(3.17). But at large $|p|$, because the last spline does not  decrease exactly as
$1/|p|$,  basis functions  very nearly, but do not exactly, satisfy the
boundary condition.  Solutions can be obtained that obey the boundary condition exactly
at large $|p|$  by extending momentum knots beyond $|p|=\infty$ \cite{Mainland:03b}
just as they were satisfied exactly by extending momentum knots below $|p|=0$. Because 
solutions decrease so rapidly at large momenta,  the value of solutions at very large
$|p|$ has minimal impact on numerical solutions. As a consequence, for a given number
of splines, more accurate solutions are obtained  without using a knot
structure that extends beyond $|p|=\infty$ and has fewer splines at small $|p|$ where
the solution has most of its support.

To solve (3.8), the dependence on $\theta_2$ and $\phi$ is first separated. Then the
resulting equation is discretized using a hybrid method:  The angular dependence is
discretized using the method of orthogonal polynomials \cite{Mainland:03a}, which
requires that the coefficient vanish independently for each of  the first  $N_\theta$
spherical functions $P^{(2)}_{\ell+I_{\theta -1},\ell}(z)$, $I_\theta=1,\dots, N_\theta$
in the equation.  The product of  functions that appear in the equation and
spherical functions that appear in the expansion for solutions can be reexpressed as
spherical functions, some of which have a larger first index. As a consequence, although
there are $N_\theta$ different spherical functions in the expansion for the solution,
there are more than $N_\theta$ different spherical functions in the equation.
Consequently, if a solution is to be obtained, the series must converge . Using the
orthogonality relationship for the spherical functions $P^{(s)}_{i,j}$,
\begin{equation}
\int^1_{-1}\mathrm{d}z(1-z^2)^{s-1 \over 2}P^{(s)}_{i,j}(z)P^{(s)}_{i',j}(z)
={\pi
\Gamma(i+j+s)
\over 2^{s-2}(2i+s) \Gamma(i-j+1) \Gamma^2(s/2)} \delta_{i,i'},
\end{equation}
it follows that multiplying the equation by 
$\sqrt{1-z^2}\;P^{(2)}_{\ell+I_{\theta-1},\ell}(z)$ and integrating over $z$ achieves
the desired discretization.  The momentum dependence is discretized using a modified
Rayleigh-Ritz-Galerkin method \cite{Delves:74, Atkinson:76}.  Thus (3.8) is converted
into a  generalized matrix eigenvalue equation of the form $Ag={\lambda
\over  m^2}Bg$, where the elements of the column vector $g$ are the expansion
coefficients $g_{n,k}$ in (3.6), by  multiplying (3.8)
by
\begin{equation}
|p|^{\mathcal N}\sqrt{1-z^2}\;{\mathcal
G}_{\ell}(|p|)B_{I_p}(|p|)P^{(2)}_{\ell+I_{\theta -1},\ell}(z),
\end{equation}
and integrating over $z$ and $|p|$.  

The expressions for the matrices $A_{i,j}$ and
$B_{i,j}$ are, respectively, 
\alphaeqn 
\begin{align}
&A_{i,j}=\int_{-1}^{1}\mathrm{d}z\sqrt{1-z^2}\int_{0}^{\infty}
\mathrm{d}|p|\,|p|^{\mathcal N} {\mathcal G}_{\ell}(|p|)\,B_{I_p}(|p|)\,
P^{(2)}_{\ell+I_{\theta}-1,\ell}(z)
\nonumber\\
&\times[D_{R}\,P^{(2)}_{\ell+J_{\theta}-1,\ell}(z)+D_{I}\,P^{(2)}_{\ell+J_{\theta}-1,
\ell}(-z)]B_{J_p}(|p|)\,{\mathcal G}_{\ell}(|p|),\\
\hspace{-1.0 cm}\mathrm{and}\nonumber\\
&B_{i,j}=2\int_{-1}^{1}\mathrm{d}z\sqrt{1-z^2}\int_{0}^{\infty}
\mathrm{d}|p|\,|p|^{({\mathcal N}-1)}\int_{0}^{\infty}\mathrm{d}|q|\,|q|^2\,\nonumber\\ 
& \times {\mathcal G}_{\ell}(|p|)\,B_{I_p}(|p|)\,
P^{(2)}_{\ell+I_{\theta}-1,\ell}(z){R(|p|,|q|)^{\ell+J_{\theta}}
\over \ell+J_{\theta}} P^{(2)}_{\ell+J_{\theta}-1,\ell}(z)\,B_{J_p}(|q|)\,{\mathcal
G}_{\ell}(|q|).
\end{align}
\reseteqn
In (3.24b)
\begin{equation}
R(|p|,|q|)=\left \{
\begin{array}{c}
\frac{|q|}{|p|}\hspace{.3 cm} \mathrm{if}\hspace{.3 cm} |q|\leq |p|\\
\\
\frac{|p|}{|q|}\hspace{.3 cm} \mathrm{if}\hspace{.3 cm} |p|\leq |q|.\\
\end{array}
\right. 
\end{equation}
As compared with (3.6), indices have been changed in (3.23) and (3.24) so that terms
are automatically excluded when $i<j$ in $P^{(2)}_{i,j}$. Here $I_p=1,
\dots,N_p$;
$I_{\theta}=1,\dots,N_{\theta}$ and the index $i$ is given by
$i=N_p(I_{\theta}-1)+I_p$ with a corresponding expression for
$j$. 

With the aid of the orthogonality relationship (3.22) for spherical functions, the
integral over the variable
$z$ in (3.24b) can be performed analytically yielding
\begin{align}
B_{i,j}=&{\pi(2\ell +I_\theta)! \over(\ell+I_\theta)(I_\theta-1)!}\int_{0}^{\infty}
\mathrm{d}|p|\,|p|^{({\mathcal N}-1)}\int_{0}^{\infty}\mathrm{d}|q|\,|q|^2\nonumber\\
&\times{\mathcal
G}_{\ell}(|p|)\,B_{I_p}(|p|) {R(|p|,|q|)^{\ell+I_\theta} \over \ell+I_\theta}
B_{J_p}(|q|)\,{\mathcal G}_{\ell}(|q|)\; \delta_{I_\theta , J_\theta}.
\end{align}

As can be seen from (3.26), if $ {\mathcal N} =3$ the matrix $B$ is both
symmetric and positive definite. Also, the matrix $A$ is  symmetric when the quantity
$D_I$ vanishes, which, from (2.5b),  occurs either when the energy is zero or when the
masses of the constituents are equal.  But when $A$ and $B$ are both symmetric and at
least one is positive definite, all eigenvalues are real \cite{Hildebrand:65}, so all
eigenvalues are real for the two cases just mentioned.  However, if the energy is
finite and the masses are unequal, all eigenvalues of the discretized equation are not
real, but real eigenvalues are obtained.  For the solutions of the discretized equation
corresponding to the lowest six real eigenvalues, which were the only solutions
checked,  when a sufficient number of basis functions were used, the solutions to the
discretized equation also satisfied the partially separated Bethe-Salpeter equation.

The disadvantage of choosing ${\mathcal N}=3$ is that the generalized matrix eigenvalue
equation actually represents the partially separated equation multiplied by
$|p|^3$.  The factor of $|p|^3$ reduces the sensitivity of the matrix equation to
the form of the solutions at small $|p|$ with the result that numerical solutions do
not satisfy the partially separated equation as well at small $|p|$. Choosing 
${\mathcal N}=1$ allows accurate solutions to be calculated for small $|p|$.  But then,
even when the energy is zero or when the masses of the constituents are equal, all
eigenvalues are not real. Nevertheless,  the same set of solutions is obtained
when ${\mathcal N}=1$ as when ${\mathcal N}=3$. Because the solutions are  more accurate
at small $|p|$ when  ${\mathcal N}=1$, all solutions in Tables 1 and 2 are calculated
with this value.  

The solutions obtained from the generalized matrix eigenvalue equation $Ag={\lambda
\over  m^2}Bg$ are checked in two ways: (1) As the number of basis functions $G_n(|p|)$
and  $P^{(2)}_{i,j}(cos\,\theta_1)$   are increased, the value of each
eigenvalue must converge.  (2) For each solution the left- and right-hand sides of
(3.7) are compared at the center of each rectangle in the physical region of the grid
formed  by the angular knots and the momentum knots. By examining where the left- and
right-hand sides of the equation agree least well, deficiencies are revealed and
possible remedies can be efficiently tested. In addition, a reliability coefficient
$r_{lhs-rhs}$ \cite{Winer:62}, which is a statistical measure of how closely the left-
and right-hand sides agree at the $(N_p)\times (N_\theta+3)$ points, is calculated. If
the left- and right-hand sides agree exactly at every point, then $r_{lhs-rhs}=1$.

Table 1 lists  values of the coupling constant $\lambda /m^2$ that are calculated
in the zero-energy limit ($\epsilon=0$) when $m_1 = 4m_2$.  Since the angular
dependence separates in the zero-energy limit, only one angular basis  function  
$P^{(2)}_{k,\ell}(cos\,\theta_1)$ is used ($N_{\theta}=1$). That single
angular basis function  is indicated by the value of the index
$k=\ell$ in the sum  (3.6).  As the number $N_p$ of momentum basis functions
$G_n(|p|)$ is increased, the calculated values of the coupling constants converge to 
correct values, and the reliability coefficients $r_{lhs-rhs}$ approach unity. (The
``exact'' eigenvalues  are correct to at least four significant figures and are
calculated numerically from the completely the separated equation \cite{Cutkosky:54}.)

\noindent Table 1.  Calculated values for the coupling constant  $\lambda /m^2$ in the
zero-energy limit when $m_1 = 4m_2$.

\begin{center}

\noindent\begin{tabular}{|c|c|c|c|c|c|c|c|}
\hline
\multicolumn{2}{|}{|c|}&\multicolumn{2}{|c|}{$N_p=5$}
&\multicolumn{2}{|c|}{$N_p=10$}&\multicolumn{2}{|c|}{$N_p=20$}\\
\hline
$\lambda /m^2_{\rm exact }$ &$\ell$&$\lambda /m^2_{\rm {calc}}$&$r_{lhs-rhs}$&$\lambda
/m^2_{\rm {calc}}$&$r_{lhs-rhs}$&$\lambda /m^2_{\rm {calc}}$&$r_{lhs-rhs}$\\
\hline
\hline
1.838&0&1.905 &0.9994&1.841&0.999990&1.838&0.99999968\\
\hline
5.000&0&5.775&0.9990&5.035&0.999993&5.000&0.99999972\\
\hline
5.654&1&5.753&0.9989&5.647&0.999996&5.652&0.99999995\\
\hline
9.817&0&11.53&0.9993&9.996&0.999993&9.822&0.99999957\\
\hline
10.43&1&11.71&0.9968&10.42&0.999991&10.42&0.99999989\\
\hline
11.46&2&11.51&0.9988&11.43&0.999986&11.45&0.99999982\\
\hline
\end{tabular}

\end{center}

Table 2 lists  values of the coupling constant $\lambda /m^2$ that are calculated
for four values of the square of the normalized energy $\epsilon^2 \equiv [E/(m_{1} +
m_{2})]^2 = 0.1, 0.5, 0.9$ and $0.99$ when $m_1 = 4m_2$.   For each energy, the number
$N_p$ of momentum basis functions  and the number $N_{\theta}$ of angular basis
functions used in the calculation are listed. As can be seen from Table 2,  as the
normalized energy $\epsilon$ increases from zero to unity (and the binding energy
decreases to zero), even using additional basis functions  it becomes increasingly
difficult to obtain  accurate eigenvalues. Nevertheless, when $\epsilon^2=0.99$, the
first eigenvalue is readily calculated with a relative error of a few tenths of a
percent, and the first six eigenvalues are all determined with relative errors less
than five percent.

\noindent Table 2.  Calculated values for the coupling constant  $\lambda /m^2$ 
when the energy is finite and  $m_1 = 4m_2$.

\begin{center}

\noindent\begin{tabular}{ll}
\noindent\begin{tabular}{|c|c|c|}
\hline
\multicolumn{3}{|c|}{$\epsilon^2 = 0.1\hspace{.5 cm}N_p=20\hspace{.5
cm}N_{\theta}=10$}\\
\hline
$\lambda /m^2_{\rm {exact}}$ &$\lambda /m^2_{\rm {calc}}$&$r_{lhs-rhs}$\\
\hline
\hline
1.686&1.686&0.99999973\\
\hline
4.690&4.691&0.99999975\\
\hline
5.156&5.154&0.99999987\\
\hline
9.252&9.264&0.99999954\\
\hline
9.688&9.669&0.99999960\\
\hline
10.42&10.41&0.9999973\\
\hline
\end{tabular}&

\begin{tabular}{|c|c|c|}
\hline
\multicolumn{3}{|c|}{$\epsilon^2 = 0.5\hspace{.5
cm}N_p=20\hspace{.5 cm}N_{\theta}=10$}\\
\hline
$\lambda /m^2_{\rm {exact}}$ &$\lambda /m^2_{\rm {calc}}$&$r_{lhs-rhs}$\\
\hline
\hline
1.052&1.052&0.9999985\\
\hline
3.112&3.111&0.9999966\\
\hline
3.344&3.341&0.9999963\\
\hline
6.174&6.185&0.9999880\\
\hline
6.532&6.493&0.9999865\\
\hline
6.748&6.757&0.9999858\\
\hline
\end{tabular}
\end{tabular}

\vspace{.5 cm}

\noindent\begin{tabular}{ll}
\noindent\begin{tabular}{|c|c|c|}
\hline
\multicolumn{3}{|c|}{$\epsilon^2 = 0.9\hspace{.5 cm}N_p=25\hspace{.5
cm}N_{\theta}=20$}\\
\hline
$\lambda /m^2_{\rm {exact}}$ &$\lambda /m^2_{\rm {calc}}$&$r_{lhs-rhs}$\\
\hline
\hline
0.3167&0.3165&0.99985\\
\hline
0.8500&0.8487&0.99973\\
\hline
1.550&1.547&0.99976\\
\hline
1.590&1.586&0.99942\\
\hline
2.534&2.522&0.99899\\
\hline
2.604&2.595&0.99918\\
\hline
\end{tabular}&

\noindent\begin{tabular}{|c|c|c|}
\hline
\multicolumn{3}{|c|}{$\epsilon^2 = 0.99\hspace{.5 cm}N_p=30\hspace{.5
cm}N_{\theta}=30$}\\
\hline
$\lambda /m^2_{\rm {correct}}$ &$\lambda /m^2_{\rm {calc}}$&$r_{lhs-rhs}$\\
\hline
\hline
0.0702&0.0700&0.968\\
\hline
0.166&0.164&0.968\\
\hline
0.286&0.273&0.954\\
\hline
0.427&0.415&0.928\\
\hline
0.590&0.613&0.871\\
\hline
0.734&0.718&0.988\\
\hline
\end{tabular}
\end{tabular}

\end{center}

\section{ Conclusions}
\numberwithin{equation}{section}
\renewcommand{\theequation}{\arabic{section}.\arabic{equation}}

A  systematic method is discussed for solving finite-energy, two-body, bound-state
Bethe-Salpeter equations that does not require that the equation be completely
separated or that the constituents have equal masses.  To apply the method, an equation
must first be Wick-rotated \cite{Wick:54} and then the two angular variables associated
with rotations in three-dimensional space must be separated, which is possible for many
two-body, bound-state Bethe-Salpeter equations, including  all such equations in the
ladder approximation. Zero-energy solutions are  calculated first:  The zero-energy
equation is completely separated by expressing the solution as a product of  a
hyperspherical harmonic and a function $F(|p|)$ that depends only on the magnitude
$|p|=(p_0^2+{\bf p}^2)^{1/2}$ of  the Euclidean four-momentum.  The zero-energy
solutions are then calculated by first expanding the function $F(|p|)$ in terms of
basis functions that (very nearly) obey the boundary conditions and discretizing
the equation by converting it into a generalized matrix eigenvalue equation that is
solved numerically. It is important to calculate zero-energy solutions first because
the basis functions that yield zero-energy solutions provide a guide for determining
the basis functions that yield finite-energy solutions. Finite-energy solutions are
calculated by expanding solutions in terms of 
basis functions, each of which is a product of a ``convergence function'' that
typically obeys the boundary conditions, a hyperspherical harmonic in four-dimensional,
Euclidean space and a spline that depends on the magnitude of the four-dimensional,
Euclidean momentum. The basis functions that yield finite-energy solutions must devolve
to the basis functions that yield zero-energy solutions in the zero-energy limit. The
partially separated equation is then discretized and solved numerically by converting
it into a generalized matrix eigenvalue equation. The generalized matrix eigenvalue
equation that yields zero-energy solutions provides guidance in formulating a
generalized matrix eigenvalue equation that yields finite-energy solutions, and the
latter must devolve to the former in the zero-energy limit.  Even though the coupling
constants, which are calculated as eigenvalues of the generalized matrix eigenvalue
equation, usually cannot all be forced to be real, real eigenvalues and corresponding
solutions are obtained that satisfy the Bethe-Salpeter equation. 

To demonstrate the  techniques and utility of the method, when the constituents have
unequal masses,  finite- and zero-energy   solutions are calculated to the partially
separated Bethe-Salpeter equation describing the Wick-Cutkosky model \cite{Wick:54,
Cutkosky:54}. For this particular equation it is convenient, but not essential, to
discretize the  angular dependence using the method of orthogonal polynomials
\cite{Mainland:03a}  and the momentum dependence  using a modified
Rayleigh-Ritz-Galerkin method \cite{Delves:74, Atkinson:76}. The advantage of
demonstrating the techniques by solving the Wick-Cutkosky model is that  complications
associated with higher spin are avoided.

Using the numerical techniques presented in the paper, the author has begun obtaining
finite-energy solutions to the scalar electrodynamics model \cite{Mainland:03a,
Mainland:03b, Sugano:56} and to the  scalar-scalar
model\cite{Schwartz:65, Nieuwenhuis:96, Kaufmann:69, Seto:93} when the bound
constituents have either equal or unequal masses.  Thus, it is highly likely that the
numerical method  discussed here provides a means for obtaining general,  finite-energy
solutions to many two-body, bound-state Bethe-Salpeter equations.

\bibliographystyle{ieeetr}
\bibliography{solving}
\end{document}